\documentclass[conference]{IEEEtran}

\IEEEoverridecommandlockouts
\usepackage{cite}
\usepackage{amsmath,amssymb,amsfonts}
\usepackage{algorithmic}
\usepackage{graphicx}
\usepackage{textcomp}
\usepackage{xcolor}
\usepackage{booktabs} 
\usepackage{multirow} 
\usepackage{subfigure}
\usepackage{caption}
\usepackage{balance}
\usepackage{marvosym}
\def\BibTeX{{\rm B\kern-.05em{\sc i\kern-.025em b}\kern-.08em
    T\kern-.1667em\lower.7ex\hbox{E}\kern-.125emX}}
\begin{document}

\title{Enhancing Attributed Graph Networks with Alignment and Uniformity Constraints for Session-based Recommendation\\
}

\author{
    \IEEEauthorblockN{
        Xinping Zhao$^{1}$, 
        Chaochao Chen$^{2}$\textsuperscript{\Letter}, \thanks{\textsuperscript{\Letter}Corresponding author.}
        Jiajie Su$^{2}$, 
        Yizhao Zhang$^{2}$, 
        Baotian Hu$^{3}$
    }
    \IEEEauthorblockA{$^{1}$School of Software Technology, Zhejiang University, Hangzhou, China}
    \IEEEauthorblockA{$^{2}$College of Computer Science and Technology, Zhejiang University, Hangzhou, China}
    \IEEEauthorblockA{$^{3}$School of Computer Science and Technology, Harbin Institute of Technology (Shenzhen), Shenzhen, China \\
        Email: \{zhaoxinping, zjuccc, sujiajie, 22221337\}@zju.edu.cn, hubaotian@hit.edu.cn}
}

\maketitle

\thispagestyle{plain} 

\begin{abstract}
Session-based Recommendation (SBR), seeking to predict a user's next action based on an anonymous session, has drawn increasing attention for its practicability. Most SBR models only rely on the contextual transitions within a short session to learn item representations while neglecting additional valuable knowledge. As such, their model capacity is largely limited by the data sparsity issue caused by short sessions. 
A few studies have exploited the Modeling of Item Attributes (MIA) to enrich item representations. However, they usually involve specific model designs that can hardly transfer to existing attribute-agnostic SBR models and thus lack universality. In this paper, we propose a model-agnostic framework, named AttrGAU ({\underline{Attr}ibuted \underline{G}raph Networks with \underline{A}lignment and \underline{U}niformity Constraints}), to bring the MIA's superiority into existing attribute-agnostic models, to improve their accuracy and robustness for recommendation.
Specifically, we first build a bipartite attributed graph and design an attribute-aware graph convolution to exploit the rich attribute semantics hidden in the heterogeneous item-attribute relationship. We then decouple existing attribute-agnostic SBR models into the graph neural network and attention readout sub-modules to satisfy the non-intrusive requirement. Lastly, we design two representation constraints, \textit{i.e.,} \textit{alignment} and \textit{uniformity}, to optimize distribution discrepancy in representation between the attribute semantics and collaborative semantics. Extensive experiments on three public benchmark datasets demonstrate that the proposed AttrGAU framework can significantly enhance backbone models' recommendation performance and robustness against data sparsity and data noise issues.
Our implementation codes will be available at https://github.com/ItsukiFujii/AttrGAU.
\end{abstract}
\section{Introduction \label{section1}}
Recommendation System (RS) plays a key role in assisting users to discover their desired items from a vast catalog of items. Conventional RS \cite{sarwar2001item,koren2009matrix} usually relies on user profiles and long-term behavior sequences. However, in many real-world scenarios, user profiles and rich behaviors are not available, due to the non-logged-in nature.
As such, Session-based Recommendation (SBR) has become a prevailing recommendation paradigm in recent years, which demonstrates promising capabilities in predicting the user's next interacted item based on a short anonymous user behavior sequence within the current session \cite{hidasi2015session,tan2016improved,li2017neural,liu2018stamp}. 
Currently, SBR has evolved to update item embeddings by constructing graph structures and then generating the session embedding by weighing different items, mainly inspired by the Graph Neural Networks (GNNs) and attention mechanisms. These GNN-based methods have achieved state-of-the-art performance in the realm of SBR \cite{wu2019session,xu2019graph,yu2020tagnn}, due to their strong capabilities in exploiting multi-hop neighbors and the significance of each item in a session.
Despite effectiveness, we argue their full potential is limited by the data sparsity issue caused by short sessions. Therefore, only mining the contextual transitions within a short session to generate the user preference has encountered a bottleneck. 
\begin{table}[t]
  \centering
  \renewcommand\arraystretch{1.25}
  \caption{The mean reciprocal rank between the target item and previously interacted ones \textit{w.r.t.} parent and leaf attributes.}
  \begin{tabular}{l|ccc}
    \toprule
   \multicolumn{1}{c|}{\multirow{1}{*}{\centering \bf Datasets}}  & \textbf{Dressipi} & \textbf{Diginetica} & \textbf{Retailrocket}\\
    \midrule 
    Parent Attribute & 83.41 & 100.0 &  42.68\\
    Leaf Attribute & 57.88 & 81.91 &  35.95\\
    \bottomrule
  \end{tabular}
  \label{tab:validation}
\end{table}
In fact, apart from interaction data, RS also involves a diverse range of exogenous data, especially the attributes of the item, which can be incorporated to model user preference more accurately \cite{xie2022decoupled,zheng2022finding,liu2021noninvasive,huang2020biane}. Particularly, we found users prefer items with the same or related attributes as those they have previously interacted with, known as \textit{Preference Similarity} \cite{wang2012recommendation}. 
To verify this claim, we conduct an experiment measuring the distance between the target (next) item and the nearest item with the same attribute, where the attribute can be the parent (\textit{e.g.,} Genre) or leaf attribute (\textit{e.g.,} Comedy, Drama, and Sci-Fi)\footnote{In real-world scenarios, attributes usually exhibit a dual-layered structure.}. Specifically, we compute the Mean Reciprocal Rank (MRR) \textit{w.r.t.} parent and leaf attributes on three public datasets (\textit{cf.} Section \ref{sec:datasets}), where we treat sessions as a ranked list from the latest clicked item to the earliest one. 
As shown in Table \ref{tab:validation}, the results fully demonstrate the above claim. For example, the results on the Dressipi dataset manifest that the latest clicked item and the latest two ones generally share the same parent or leaf attributes as the target item, respectively. These observations show that effectively modeling item attributes has great potential to mitigate the data sparsity issue.
However, the Modeling of Item Attributes (MIA) does not receive much attention in the literature, and we believe this factor plays a key role in learning user preferences according to the above analysis, especially when facing the severe data sparsity issue\footnote{In our experiments, we found that the sparser the training data, the greater the benefit by incorporating the modeling of item attributes, \textit{cf.} Section \ref{robust_sparse}.}. 
Existing attribute-aware SBR approaches typically involve specific model designs \cite{lai2022attribute,ma2023clhhn,xu2022category} and lack universality, whose techniques do not apply to other attribute-agnostic models. Here we wish to bring the MIA's superiority into existing attribute-agnostic models by developing a general model-agnostic framework, which meets the non-intrusive requirement and offers flexible usability. To achieve these goals, there remain two challenges that need to be addressed:
\begin{itemize}
    \item \textbf{Heterogeneous Item-Attribute Relationship.} The relationship between items and attributes exhibits highly heterogeneous properties. More specifically, the types and quantities of attributes associated with each item vary significantly. For example, a mobile phone typically involves attributes such as processor type, screen size, and brand, whereas a piece of clothing commonly involves attributes like style, color, and so on. Therefore, how to efficiently organize such intricate structures and extract informative\, semantics\, from\, them\, becomes\, a\, challenge.    
    \item \textbf{Distribution Discrepancy in Representation.} Due to the large semantic gap between the \textit{attribute semantics} and \textit{collaborative semantics} \cite{zheng2023adapting}, there exists a considerable distribution discrepancy between the raw and attribute-enriched item representations, which would impair the model's recommendation performance. Moreover, the GNN and attention sub-modules in existing SBR approaches would enlarge the impact of distribution discrepancy on representation learning, making the representation learning less robust to the semantic gap. Therefore, how to refine the session representations from the view of distribution discrepancy becomes a challenge.
\end{itemize}
In this paper, we propose a model-agnostic framework, named {AttrGAU} 
({\underline{Attr}ibuted \underline{G}raph Networks with \underline{A}lignment and \underline{U}niformity constraints}), to enhance the performance of existing attribute-agnostic SBR approaches.
Specifically, it comprises three key modules: \textit{{attribute-aware graph modeling}}, \textit{session representation learning}, and \textit{alignment\&uniformity constraints}.
\textbf{(i)} In the \textit{{attribute-aware graph modeling}}, we first organize the item-attr\footnote{In the pages that follow, we employ `attr' to denote the attribute for brevity.} data as a Bipartite Attributed Graph (BAG), with the items and leaf attrs represented as the node, and the parent attrs represented as the edges so that the heterogeneous item-attr relationship is well preserved.
Then, we propose an attribute-aware graph convolution to refine the node embeddings via aggregating informative features from their neighbors. Lastly, considering the over-smoothing issue, we design a node-level cross-layer contrast regularization to enforce node differences. 
\textbf{(ii)} In the \textit{{session representation learning}}, we decouple the existing SBR approaches into two plug-and-play components, \textit{i.e.,} the graph neural network sub-module to update node embeddings and the attention readout sub-module to generate the session embedding.
Obeying the non-intrusive requirement, we use dual embedding for items to represent and propagate the raw and processed information separately, to capture the holistic semantics of items better. Lastly, a fused session embedding is used to make the final recommendation. 
\textbf{(iii)} Additionally, we introduce two representation constraints:  \textit{alignment} and \textit{uniformity}, to optimize distribution discrepancy in representations. The alignment constraint forces the representations from the same session to be as close as possible. 
However, if only alignment is considered, perfectly aligned encoders are easy to achieve by mapping all the session embeddings to the same representation. To avoid this problem, the uniformity constraint forces the representations from the different sessions to be as distant as possible. 
Our \textbf{main contributions} can be summarized as the following three-fold: \textbf{(1) Idea.} To the best of our knowledge, our study is the first to explore a general solution to enhance existing attribute-agnostic SBR approaches via integrating item attribute modeling. 
\textbf{(2) Methodology.} We propose a model-agnostic framework, {AttrGAU}, which can effectively deal with heterogeneous item-attr relationships and optimize distribution discrepancy in representations. Furthermore, {AttrGAU} satisfies the two properties of being non-intrusive and flexible for plug-and-play usage. 
\textbf{(3) Experiment.} We conduct extensive experiments on three public benchmark datasets. The experimental results show that AttrGAU can significantly enhance the existing attribute-agnostic SBR models' performance and endow them with more robustness to the data sparsity issues. 
\section{Preliminaries}
\subsection{Problem Statement \label{Problem_Statement}}
In a typical SBR scenario, we have an item set $\mathcal{V}=\{v_1, v_2, ..., v_{|\mathcal{V}|}\}$, where $|\mathcal{V}|$ is the total number of items. 
Each anonymous session, which can be denoted as $s = \{v_{s,1}, v_{s,2}, ...,v_{s,n}\}$, consists of a sequence of interactions (\textit{e.g.,} clicks and views) in chronological order, where $v_{s, i} \in \mathcal{V}$ represents an interacted item of the user at the $i$-th timestamp. 
In real-world scenarios, an item usually contains multiple dual-layered attributes. For example, a movie could contain parent attributes such as Genre and Language, where each of its parent attributes is associated with a leaf attribute such as Comedy and English. 
Formally, we denote $\mathcal{P} = \{p_1, p_2, ..., p_{|\mathcal{P}|}\}$ as the parent attribute set and $\mathcal{Q} = \{q_1, q_2, ..., q_{|\mathcal{Q}|}\}$ as the leaf attribute set. Given the entire attribute information, we organize it by an item-parent-leaf\footnote{The `parent' and `leaf' are the abbr of the parent and leaf attr, respectively.} incidence tensor $\mathbf{X} \in \mathbb{R}^{(|\mathcal{V}|+|\mathcal{P}|+|\mathcal{Q}|) \times (|\mathcal{V}|+|\mathcal{P}|+|\mathcal{Q}|)}$, where each nonzero entry ($i, c, a$) denotes that item $v_i$ has parent attribute $p_c$ and its leaf attribute $q_a$. Furthermore, we utilize $\mathbf{R} \in \mathbb{R}^{|\mathcal{V}|\times|\mathcal{P}|}$, $\mathbf{H} \in \mathbb{R}^{|\mathcal{V}|\times|\mathcal{Q}|}$, and $\mathbf{B} \in \mathbb{R}^{|\mathcal{P}|\times|\mathcal{Q}|}$ to denote item-parent incidence matrix, item-leaf incidence matrix, and parent-leaf cooccurrence matrix, respectively. Notably, the value in $\mathbf{R}$ and $\mathbf{H}$ may be greater than one since an item could contain multiple attributes of the same type.
Given the incidence tensor $\mathbf{X}$, we convert it to the form of the bipartite attributed graph, where each vertex represents an item or a leaf attr, and each edge represents a connection between the item and leaf attr under a parent attr. Then, the neighbors of the item $v_i$ can be denoted as $\mathcal{N}_i = \{(a,c)|\mathbf{X}_{i,c,a}=1\}$, and the neighbors of the leaf attr $q_a$ can be denoted as $\mathcal{N}_a = \{(i,c)|\mathbf{X}_{i,c,a}=1\}$. 
Based on the above definitions, we can define the task of attribute-aware SBR. Specifically, given the previous interaction sequence $s$ as well as the item-attr association information $\mathbf{X}$, it focuses on predicting the next most possible item $v_{s,n+1}$, which can be formulated as:
\begin{equation}
    v_{*} = \arg\max_{v_i\in\mathcal{V}} P(v_{s,n+1}=v_i|s,\mathbf{X}).
\end{equation}
\begin{figure}[t]
    \centering
    \includegraphics[width=1.\linewidth]{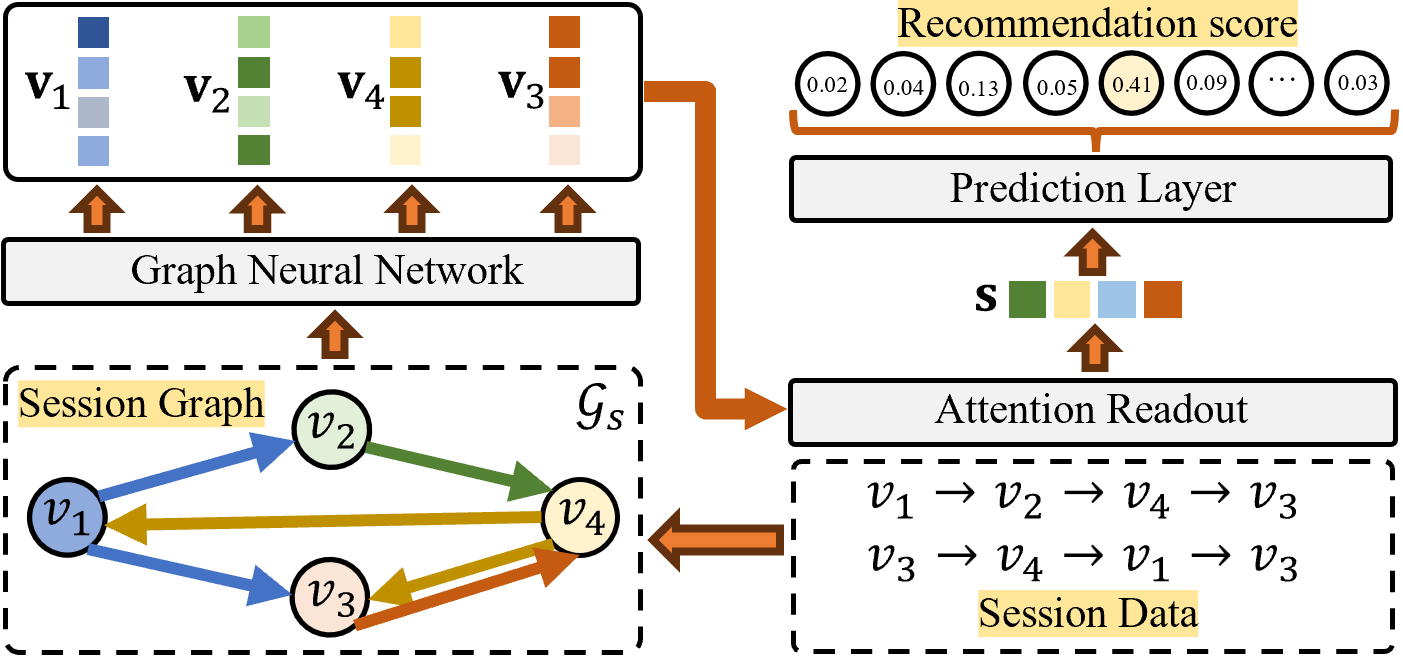}
    \caption{The backbone structure of GNN-based SBR models.}
    \label{fig:backbone_framework}
\end{figure}
\subsection{Backbones for SBR \label{sec:backbone}}
Graph neural networks (GNN) are a powerful way to encode sessions, leading to state-of-the-art performance in SBR. Therefore, we choose three representative GNN-based SBR models as backbones, \textit{i.e.,} SR-GNN \cite{wu2019session}, GC-SAN \cite{xu2019graph}, and TAGNN \cite{yu2020tagnn}. They comprise two key components: the \textit{\textbf{graph neural network sub-module}} and \textit{\textbf{attention readout sub-module}}. 
We illustrate the backbone structure of SBR in Figure \ref{fig:backbone_framework}. The graph neural network sub-module first represents each session $s$ as a session graph $\mathcal{G}_s = (\mathcal{V}_s, \mathcal{E}_s, \mathbf{A}_s)$, where $\mathcal{V}_s, \mathcal{E}_s, \mathbf{A}_s$ are the node set, the edge set, and the adjacency matrix, respectively. After that, it updates each node embedding in a session graph $\mathcal{G}_s$ by aggregating and combining the embeddings of their neighbors, which\, can\, be\, formulated\, as:
\begin{equation}
    \left\{\mathbf{v}_{i}\right\}_{i=1}^{n} = \mathrm{H}_{\mathrm{gnn}}\left(\{\mathbf{v}_{i}^{(0)} | i = {1,2,... ,n} \}, \mathcal{G}_{s}\right),
    \label{Equ:backbone_gnn}
\end{equation}
where $\mathbf{v}_{i},\mathbf{v}_{i}^{(0)} \in \mathbb{R}^d$ denote the encoded and raw embedding for item $v_{s,i}$, respectively; $\mathrm{H}_{\mathrm{gnn}}(\cdot)$ represents the neighborhood aggregation and combination function, such as GGNN \cite{li2015gated} employed in SR-GNN. 
The attention readout sub-module further utilizes the attention mechanism to model the significance of different items in a session, and then generate \quad \quad \quad the session representation through weighting and transforming:
\begin{equation}
    \mathbf{s} = \mathrm{H}_{\mathrm{att}}\big(\left\{\mathbf{v}_{i} | i = {1,2,... ,n} \right\}\big),
    \label{Equ:backbone_att}
\end{equation}
where $\mathbf{s} \in \mathbb{R}^d$ denotes the session representation; $\mathrm{H}_{\mathrm{att}}(\cdot)$ represents the attention readout function, such as the additive attention \cite{bahdanau2014neural} employed in SR-GNN. Subsequently, a prediction layer is built upon the session representation $\mathbf{s}$ to predict how likely $v_i$ would be the next item. The inner product is a widely used solution to compute the recommendation score $\mathbf{\hat{z}}_i$. 
Following that, the softmax function is adopted to handle the unnormalized recommendation score $\mathbf{\hat{z}}_i$ for all candidates:
\begin{equation}
    \mathbf{\hat{y}} = \mathrm{softmax}(\mathbf{{z}}), \;\; \mathbf{\hat{z}}_i = \mathbf{s}^\mathrm{T} \mathbf{v}_i^{(0)}, 
    \label{Equ:backbone_pre}
\end{equation}
where $\mathbf{\hat{y}} \in \mathbb{R}^{|\mathcal{V}|}$ denotes the probabilities of items being the next item. Finally, the recommendation learning loss is defined as\, the\, cross-entropy of\, the\, prediction\, and\, the\, ground\, truth:
\begin{equation}
    \mathcal{L}_{rec} = - \sum_{i=1}^{|\mathcal{V}|} \mathbf{{y}}_i \log(\mathbf{\hat{y}}_i) + (1-\mathbf{{y}}_i)\log(1-\mathbf{\hat{y}}_i),
    \label{Equ:backbone_rec}
\end{equation}
where $\mathbf{{y}} \in \{0, 1\}^{|\mathcal{V}|}$ denotes the one-hot vector of the ground truth item. Here, we choose it as the supervised learning task.
\section{Methodology}
\begin{figure*}[t]
    \centering
    \includegraphics[width=1.0\linewidth]{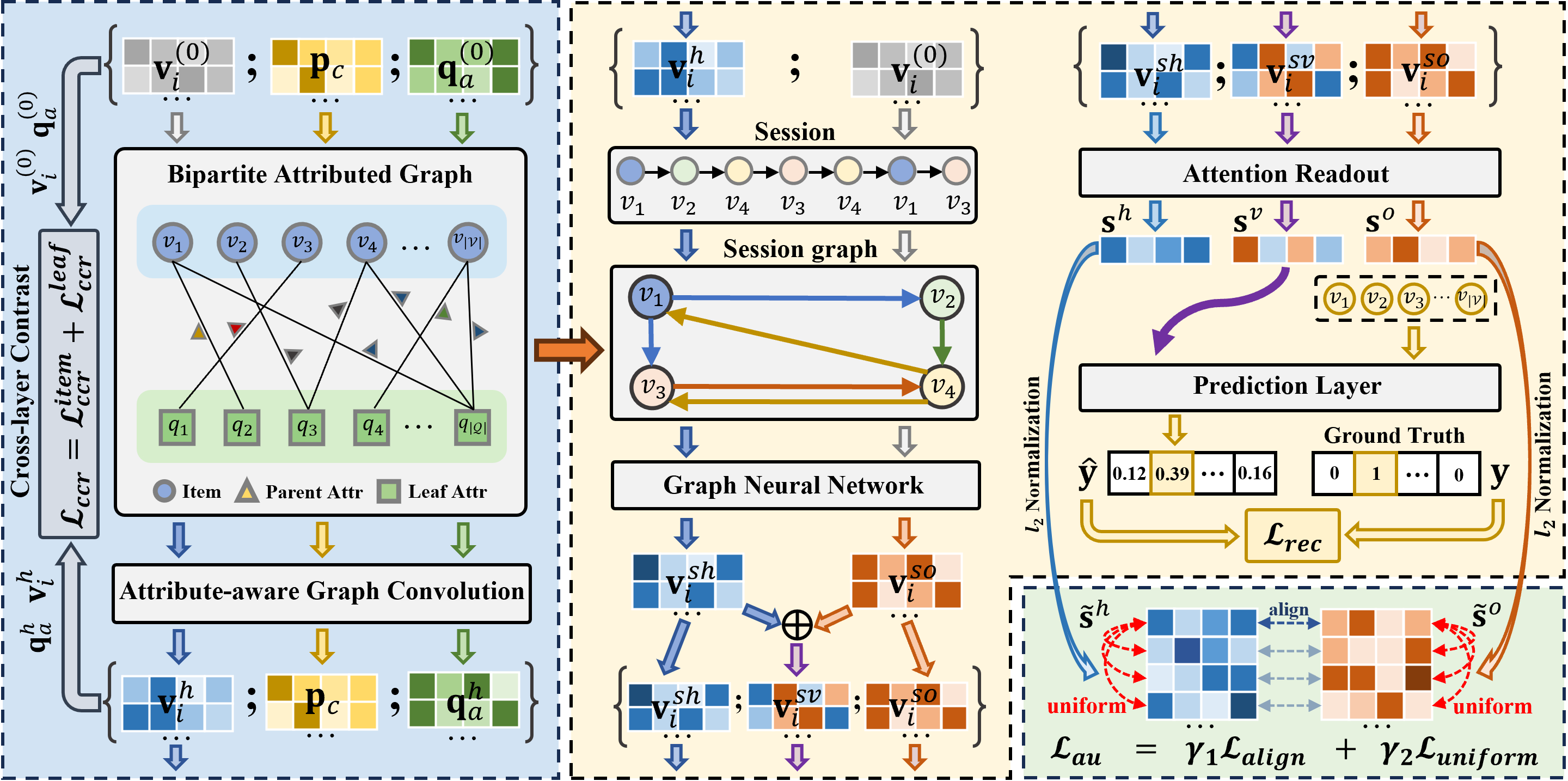}
    \caption{Overall framework of the proposed AttrGAU. The three parts show Attribute-aware Graph Modeling (marked in \textcolor{blue}{blue}), Session Representation Learning (marked in \textcolor[RGB]{194,144,0}{yellow}), and Alignment and Uniformity Constraints (marked in \textcolor[RGB]{84,130,53}{green}), respectively.}
    \label{fig:model_framework}
\end{figure*}

Figure \ref{fig:model_framework} depicts the overall framework of the proposed AttrGAU, which mainly consists of three modules: \textbf{(i)} \textbf{\textit{attribute-aware graph modeling}}, which learns attribute-enriched item representations by modeling heterogeneous item-attr patterns; \textbf{(ii)} \textbf{\textit{session representation learning}}, which utilizes a plug-and-play SBR backbone to learn session representations by capturing contextual transitions and modeling item contributions; and \textbf{(iii)} \textbf{\textit{alignment\&uniformity constraints}}, which optimize distribution discrepancy in representation via explicitly regularizing the resultant raw and attribute-enriched session representations. We introduce them at length in the following.
\subsection{Attribute-aware Graph Modeling}
The prime task of AttrGAU is to enrich the raw item representations by extracting informative semantic patterns from the heterogeneous item-attribute relationships. \textbf{\textit{1)}} To this end, we propose the \textbf{\textit{attribute-aware graph convolution}} that integrates the attributed context into neighborhood aggregation to learn attribute-enriched item representations. \textbf{\textit{2)}} Simultaneously, considering the serious over-smoothing issue in item representations, we design a \textbf{\textit{cross-layer contrast regularization}} to enforce node differences via node-level discrimination.
\subsubsection{\textbf{Attribute-aware Graph Convolution}} We first design the convolution schema of item-attr in the Bipartite Attributed Graph (BAG). As an item $v_i$ involves multiple parent-leaf attribute pairs, its neighborhood\footnote{Note that the term `neighborhood' includes both the adjacent nodes and the connecting edges.} can reflect the semantic similarity between $v_i$ and its connected parent attrs, as well as leaf attrs to a large extent. Ditto for a leaf attr $q_a$, its neighborhood can characterize the semantic feature of $q_a$ well. 
Formally, given a target item $v_i$ and a leaf attr $q_a$ in BAG, aggregating local information from their neighbors in BAG can effectively refine their raw representations, making them more robust against the data sparsity issue. Therefore, we use $\mathcal{N}_i$ and $\mathcal{N}_a$ (\textit{cf.} Section \ref{Problem_Statement}) to represent the first-order adjacent nodes and connecting edges of item $v_i$ and leaf attr $q_a$, respectively, and we propose to integrate the attributed context from neighborhood nodes as well as connecting edges to learn the refined representations of item $v_i$ and leaf attr $q_a$:
\begin{equation}
\begin{aligned}
    \mathbf{v}_i^{(l+1)} &= \mathrm{H}_{\mathrm{attrgc}}\left(\{( \mathbf{v}_i^{(l)} , \mathbf{p}_c, \mathbf{q}_a^{(l)})|(a, c) \in \mathcal{N}_i \}\right), \\
    \mathbf{q}_a^{(l+1)} &= \mathrm{H}_{\mathrm{attrgc}}\left(\{(\mathbf{q}_a^{(l)}, \mathbf{p}_c, \mathbf{v}_i^{(l)})| (i, c) \in \mathcal{N}_a \}\right),
\end{aligned}
\end{equation}
where $\mathrm{H}_{\mathrm{attrgc}}(\cdot)$ denotes the Attribute-aware Graph Convolution (AttrGC) function to extract and integrate information associated with $v_i$ and $q_a$ from their connections in BAG; $\mathbf{v}_i^{(l)},\mathbf{q}_a^{(l)} \in \mathbb{R}^d$ denotes the refined item and leaf attr representations of the $l$-th AttrGC layer, respectively; $\mathbf{p}_c \in \mathbb{R}^d$ denotes the parent attr representation; $d$ is the embedding size. 
Previous studies \cite{wu2022graph} have shown that standard Graph Convolution does not consider the features on edges, while they are important to understand the attributed context between two connected nodes. Consequently, it is necessary to integrate the connecting edges into representation learning. To this end, we propose a simple yet effective attribute-aware graph convolution operation that extracts and integrates informative patterns from\, both\, the\, adjacent\, nodes\, and\, connecting\, edges:
\begin{equation}
\begin{aligned}
    \mathbf{v}_i^{(l+1)} &= \sum_{(a, c) \in \mathcal{N}_i} \left(\frac{1}{\sqrt{|\mathcal{N}_i|}\sqrt{|\mathcal{N}_a|}} \mathbf{q}_a^{(l)} +  \frac{1}{{|\mathcal{N}_i|}}  \mathbf{p}_c \right), \\
    \mathbf{q}_a^{(l+1)} &= \sum_{(i, c) \in \mathcal{N}_a} \left(\frac{1}{\sqrt{|\mathcal{N}_a|}\sqrt{|\mathcal{N}_i|}} \mathbf{v}_i^{(l)} +  \frac{1}{{|\mathcal{N}_a|}}  \mathbf{p}_c \right),
    \label{Equ:layer-wise}
\end{aligned}
\end{equation}
where $|\mathcal{N}_i|$ and $|\mathcal{N}_a|$ denote the number of edges connected with the item $v_i$ and leaf attr $q_a$, respectively; the symmetric normalization term $\frac{1}{\sqrt{|\mathcal{N}_i|}\sqrt{|\mathcal{N}_a|}}$ (or $\frac{1}{\sqrt{|\mathcal{N}_a|}\sqrt{|\mathcal{N}_i|}}$) is used to avoid the scale of embedding values increasing with the graph convolution operations, whose effectiveness has been fully verified by NGCF \cite{wang2019neural} and LightGCN \cite{he2020lightgcn}; the $L_1$ normalization term $\frac{1}{{|\mathcal{N}_i|}}$ (or $\frac{1}{{|\mathcal{N}_a|}}$) is employed to average the features on edges because they are unsymmetrical. It is worth noting that we have also tried other types of normalization terms. However, they do not lead to performance improvement compared to the $L_1$ normalization term. 
Since different layers hold different semantics, an item (or a leaf attr) representation can be further refined by aggregating information from its multi-hop neighbors. Therefore, we stack multiple AttrGC layers and combine the representations obtained 
at each layer to form the holistic representation of an item \,(or a leaf attr):
\begin{equation}
\textstyle
    \mathbf{v}_i^{h} = \sum_{l=0}^{L} \alpha_l \mathbf{v}_i^{(l)}, \quad \mathbf{q}_a^{h} = \sum_{l=0}^{L} \alpha_l \mathbf{q}_a^{(l)},
\end{equation}
where $L$ denotes the total number of AttrGC layers; $\mathbf{v}_i^{h}, \mathbf{q}_a^{h} \in \mathbb{R}^d$ denote the holistic item and leaf attr representations, respectively; $\alpha_l>0$ denotes the importance of the $l$-th layer representation, which can be treated as a hyper-parameter and tuned manually with the constraint that $\sum_{l=0}^L \alpha_l=1$. In our experiments, we find that uniformly setting $\alpha_l$ as $\frac{1}{L+1}$ usually leads to good performance. Therefore, we do not design a special module to learn and optimize $\alpha_l$ here and leave this for future exploration because it is not the point of this work.
To speed up the calculation of the AttrGC operation, we implement it in the matrix manipulation form. Specifically, given the item-parent incidence matrix $\mathbf{R} \in \mathbb{R}^{|\mathcal{V}|\times|\mathcal{P}|}$, item-leaf
incidence matrix $\mathbf{H} \in \mathbb{R}^{|\mathcal{V}|\times|\mathcal{Q}|}$, and parent-leaf cooccurrence matrix $\mathbf{B} \in \mathbb{R}^{|\mathcal{P}|\times|\mathcal{Q}|}$, we define the item-parent-leaf adjacency matrix of the bipartite attributed graph\, as\, follows:
\begin{equation}
    \mathbf{A}  = 
        \begin{bmatrix}  
        \mathbf{0} & \mathbf{R} & \mathbf{H}\\  
        \mathbf{0} & \mathbf{I} & \mathbf{0}\\  
        \mathbf{H}^\mathrm{T} & \mathbf{B}^\mathrm{T} & \mathbf{0}\\ 
        \end{bmatrix},
\end{equation}
where $\mathbf{0}, \mathbf{I}$ are the zero matrix and the identity matrix, respectively. Subsequently, given the diagonal degree matrix of $\mathbf{A}$, we define its diagonal degree matrix as $\mathbf{D} \in \mathbb{R}^{(|\mathcal{V}|+|\mathcal{P}|+|\mathcal{Q}|) \times (|\mathcal{V}|+|\mathcal{P}|+|\mathcal{Q}|)}$, whose $k$-th diagonal element is $\mathbf{D}_{kk} = \sum_{j=1} \mathbf{A}_{kj}$. However, there are some cases in which the diagonal elements of the degree matrix are double-counting. Specifically, the parent and leaf attributes of each item are counted twice. Hence, we propose further correcting the\, diagonal\, degree\, matrix\, $\mathbf{D}$, which\, can be\, formulated\, as:
\begin{equation}
\nonumber
    \mathbf{\Tilde{D}}_{kk} =  \begin{cases}
    \frac{1}{2}\mathbf{D}_{kk},&\!\!\!\!k \in \left[0, |\mathcal{V}|\right) \cup \left[|\mathcal{V}|+|\mathcal{P}|, |\mathcal{V}|+|\mathcal{P}|+|\mathcal{Q}|\right) \\
    \mathbf{D}_{kk},&\!\!\!\!k \in \left[|\mathcal{V}|, |\mathcal{V}|+|\mathcal{P}|\right)
    \end{cases},
\end{equation}
where $\mathbf{\Tilde{D}}  \in \mathbb{R}^{(|\mathcal{V}|+|\mathcal{P}|+|\mathcal{Q}|) \times (|\mathcal{V}|+|\mathcal{P}|+|\mathcal{Q}|)}$ denotes the corrected diagonal degree matrix. We next define two mask matrices (\textit{i.e.,} $\mathbf{M}_1$ and $\mathbf{M}_2$) for processing the adjacency matrix $\mathbf{A}$ to adapt varying normalization terms needed for different components\, in\, the\, proposed\, attribute-aware\, graph\, convolution: 
\begin{equation}
    \mathbf{M}_1 = \begin{bmatrix}  
        \mathbf{0} & \mathbf{0} & \mathbf{\Tilde{H}}\\  
        \mathbf{0} & \mathbf{0} & \mathbf{0}\\  
        \mathbf{\Tilde{H}}^{\mathrm{T}} & \mathbf{0} & \mathbf{0}\\ 
        \end{bmatrix}, \quad 
    \mathbf{M}_2 = \begin{bmatrix}  
        \mathbf{0} & \mathbf{\Tilde{R}} & \mathbf{0}\\  
        \mathbf{0} & \mathbf{{I}} & \mathbf{0}\\  
        \mathbf{0} & \mathbf{\Tilde{B}}^{\mathrm{T}} & \mathbf{0}\\ 
        \end{bmatrix},
\end{equation}
where $\mathbf{\Tilde{H}}\in\{0,1\}^{|\mathcal{V}|\times|\mathcal{Q}|},\mathbf{\Tilde{R}}\in\{0,1\}^{|\mathcal{V}|\times|\mathcal{P}|}$, and $\mathbf{\Tilde{B}}\in\{0,1\}^{|\mathcal{P}|\times|\mathcal{Q}|}$ are the binarization matrix of $\mathbf{H}$, $\mathbf{R}$, and $\mathbf{B}$, respectively.\! The binarization processing can be formulated as:
\begin{footnotesize}
\begin{equation}
\nonumber
    \mathbf{\Tilde{H}}_{i,a} =  \begin{cases}
    1, &\!\!\!\! \mathbf{H}_{i,a} \neq 0\\
    0, &\!\!\!\! \mathbf{H}_{i,a} = 0
    \end{cases}, \,
    \mathbf{\Tilde{R}}_{i,c} =  \begin{cases}
    1, &\!\!\!\! \mathbf{R}_{i,c} \neq 0\\
    0, &\!\!\!\! \mathbf{R}_{i,c} = 0
    \end{cases}, \,
    \mathbf{\Tilde{B}}_{c,a} =  \begin{cases}
    1, &\!\!\!\! \mathbf{B}_{c,a} \neq 0\\
    0, &\!\!\!\! \mathbf{B}_{c,a} = 0
    \end{cases}.
\end{equation}
\end{footnotesize}Based on the above definitions (\textit{i.e.,} the adjacency matrix $\mathbf{A}$, the corrected diagonal degree matrix $\Tilde{\mathbf{D}}$, and the mask matrices), the normalized adjacency matrix can be defined as:
\begin{equation}
    \mathbf{\Tilde{A}} = \mathbf{\Tilde{D}}^{-\frac{1}{2}}\left(\mathbf{A}\mathbf{\Tilde{D}}^{-\frac{1}{2}} \odot \mathbf{M}_1 + \mathbf{\Tilde{D}}^{-\frac{1}{2}} \mathbf{A} \odot \mathbf{M}_2 \right).
\end{equation}
Then, we implement the layer-wise propagation rule via a matrix manipulation form, which can be formulated as follows:
\begin{equation}
    \mathbf{E}^{(l+1)}  = \mathbf{\Tilde{A}}\mathbf{E}^{(l)},
\end{equation}
where $\mathbf{E}^{(l)} \in \mathbb{R}^{(|\mathcal{V}|+|\mathcal{P}|+|\mathcal{Q}|) \times d}$ is the concatenation of the item, parent attr, and leaf attr embedding matrix. Particularly, $\mathbf{E}^{(0)}$ is the concatenation of their original embedding matrices: 
\begin{equation}
    \mathbf{E}^{(0)} = \mathbf{E} = \left[\mathbf{v}_1^{(0)}, ..., \mathbf{v}_{|\mathcal{V}|}^{(0)},\mathbf{p}_1, ..., \mathbf{p}_{|\mathcal{P}|},\mathbf{q}_1^{(0)}, ..., \mathbf{q}_{|\mathcal{Q}|}^{(0)}\right]^\mathrm{T}\!\!\!\!\!,
\end{equation}
where $\mathbf{v}_*^{(0)}, \mathbf{p}_*, \mathbf{q}_*^{(0)} \in \mathbb{R}^d$ are the original embedding vector of item $v_*$, parent attr $p_*$, and leaf attr $q_*$, respectively. Lastly, we get the holistic embedding matrix $\mathbf{E}^{h} \in \mathbb{R}^{(|\mathcal{V}|+|\mathcal{P}|+|\mathcal{Q}|) \times d}$ used to enhance the robustness of the existing SBR backbone models\, against\, data\, sparsity,\, which\, can\, be\, formulated\, as:
\begin{align}
 \mathbf{E}^h &= \left[\mathbf{v}_1^h, ..., \mathbf{v}_{|\mathcal{V}|}^h,\mathbf{p}_1, ..., \mathbf{p}_{|\mathcal{P}|},\mathbf{q}_1^h, ..., \mathbf{q}_{|\mathcal{Q}|}^h\right]^{\mathrm{T}}  \nonumber\\
   &= \alpha_0 \mathbf{E}^{(0)} + \alpha_1 \mathbf{E}^{(1)} + \alpha_2 \mathbf{E}^{(2)} + ... + \alpha_L \mathbf{E}^{(L)} \\
   &= \alpha_0 \mathbf{E}^{(0)} + \alpha_1  \mathrm{\Tilde{A}}\mathbf{E}^{(0)} + \alpha_2 \mathrm{\Tilde{A}}^2 \mathbf{E}^{(0)} + ... + \alpha_L  \mathrm{\Tilde{A}}^L\mathbf{E}^{(0)}. \nonumber
\end{align}
\subsubsection{\textbf{Cross-layer Contrast Regularization}}  As the number of the graph convolution layer increases, an item (or a leaf attr) representation can be further refined by aggregating the features from their multi-hop neighbors. 
However, this inevitably causes the over-smoothing issue \cite{chen2020measuring}, which makes embeddings locally similar and aggravates the Matthew Effect. Previous studies \cite{yu2023xsimgcl,DBLP:conf/iclr/Cai0XR23} have shown that contrastive learning can effectively mitigate the over-smoothing issue. 
Besides, an empirical study \cite{xue2023study} on the relationship between transformer configuration and training objectives suggests that token-level training objectives are more suitable for scaling models along depth than sequence-level ones. Inspired by the above studies, we propose to contrast the representations between the holistic and $0$-th layer representation, which avoids additional computational overhead from data augmentation and effectively mitigates the over-smoothing issue. 
Specifically, we treat the holistic and $0$-th layer representations of the same item (or leaf attr) as the positive pairs (\textit{i.e.,} $\{(\mathbf{v}_i^{h},\mathbf{v}_i^{(0)})| i \in \mathcal{V}\}$ and $\{(\mathbf{q}_a^{h},\mathbf{q}_a^{(0)})| a \in \mathcal{Q}\}$). Conversely, the holistic and $0$-th layer representations of any different items (or leaf attrs) are treated as negative pairs (\textit{i.e.,} $\{(\mathbf{v}_i^{h},\mathbf{v}_{i^-}^{(0)})| i, i^- \in \mathcal{V}, i \neq i^-\}$ and $\{(\mathbf{q}_a^{h},\mathbf{q}_{a^-}^{(0)})| a, a^- \in \mathcal{Q}, a \neq a^-\}$). 
Formally, we follow SGL \cite{wu2021self} and adopt the contrastive loss, InfoNCE \cite{gutmann2010noise}, to implement our proposed cross-layer contrast regularization by maximizing the agreement of positive pairs and minimizing the agreement of negative pairs, which can be formulated as:
\begin{align}
    \nonumber   
    \mathcal{L}_{ccr}^{item} &= \sum_{i \in \mathcal{V}} - \log \frac{e^{s(\mathbf{v}_i^h, \mathbf{v}_i^{(0)})/\tau}}{e^{s(\mathbf{v}_i^h, \mathbf{v}_{i}^{(0)})/\tau} + \sum_{i^{-} \in \mathcal{V}\backslash\{i\}} e^{s(\mathbf{v}_i^h, \mathbf{v}_{i^{-}}^{(0)})/\tau}}, \\
    \nonumber   
    \mathcal{L}_{ccr}^{leaf} &= \sum_{a \in \mathcal{Q}} - \log \frac{e^{s(\mathbf{q}_a^h, \mathbf{q}_a^{(0)})/\tau}}{e^{s(\mathbf{q}_a^h, \mathbf{q}_{a}^{(0)})/\tau}+\sum_{a^{-} \in \mathcal{Q}\backslash\{a\}} e^{s(\mathbf{q}_a^h, \mathbf{q}_{a^{-}}^{(0)})/\tau}},\\
    \mathcal{L}_{ccr} &= \mathcal{L}_{ccr}^{item}+\mathcal{L}_{ccr}^{leaf}, \label{Equ:ccr}
\end{align}
where $s(\cdot)$ is the cosine similarity function; $\tau$ is the temperature coefficient employed to control the distribution's kurtosis.
\subsection{Session Representation Learning}
Having established the attribute-enriched and raw item representations, we employ plug-and-play SBR backbone models to learn session representations. Formally, they consist of two key components: \textbf{\textit{1)}} \textit{\textbf{graph neural network sub-module}}, which exploits the complex contextual transitions among items in a session via the graph neural network, and \textbf{\textit{2)}} \textit{\textbf{attention readout sub-module}}, which models the contributions of different items in\, a\, session\, via\, the\, attention\, mechanism\, (\textit{cf.} Section \ref{sec:backbone}). 
\subsubsection{\textbf{Graph Neural Network Sub-module}} Unlike specifically designed attribute-aware SBR models \cite{lai2022attribute,ma2023clhhn,xu2022category}, we hope to bring the MIA's superiority into existing attribute-agnostic SBR backbone models while satisfying the non-intrusive requirement.
To this end, we propose exploiting contextual transitions separately in a session with dual-item representations. Specifically, we utilize the neighborhood aggregation and combination function $\mathrm{H}_{\mathrm{gnn}}(\cdot)$ to encode the attribute-enriched item representations $\{\mathbf{v}_i^h\}_{i=1}^n$ and the raw item representations $\{\mathbf{v}_i^{(0)}\}_{i=1}^n$, and then obtain the encoded representations $\{\mathbf{v}_i^{sh}\}_{i=1}^n$ and $\{\mathbf{v}_i^{so}\}_{i=1}^n$ (\textit{cf.} Equation (\ref{Equ:backbone_gnn})). 
Considering that quite a few raw item representations are semantically poor owing to data sparsity, we generate the final representation of each item by incorporating its encoded representation\! of\! different\! channels,\! which\! can\! be\! formulated\! as:
\begin{equation}
    \mathbf{v}^{sv}_{i} = \mathrm{dropout}(\mathbf{v}^{sh}_{i})+\mathrm{dropout}(\mathbf{v}^{so}_{i}),
\end{equation}
where $\mathbf{v}^{sv}_{i} \in \mathbb{R}^d$, $1 \leq i \leq n$; $\mathbf{v}^{sh}_{i}$ and $\mathbf{v}^{so}_{i}$ are the attribute-enriched and raw item representations, respectively. It is worth mentioning that $\mathbf{v}^{so}_{i}$ is the encoded item representation of mining purely contextual transitions in a session, while $\mathbf{v}^{sh}_{i}$ enhances the transition modeling with attribute semantics. 
Moreover, we further employ the dropout technique \cite{srivastava2014dropout} on the encoded\, representations\, to\, avoid\, the\, problem\, of\, overfitting. 
\subsubsection{\textbf{Attention Readout Sub-module}} In fact, the contribution of different items within a session $s$ is usually not equal \textit{w.r.t.} the next item prediction \cite{wang2020global}. Because of this, previous studies \cite{wu2019session,xu2019graph,yu2020tagnn} commonly adopt the attention mechanism to model the significance of different items in a session, and then obtain the session representation via weighting and transforming. 
Specifically, we utilize the attention readout function $\mathrm{H}_{\mathrm{att}}(\cdot)$ to model the encoded item representations $\{\mathbf{v}_i^{sh}\}_{i=1}^n$, $\{\mathbf{v}_i^{sv}\}_{i=1}^n$, and $\{\mathbf{v}_i^{so}\}_{i=1}^n$, and then generate the session representations $\mathbf{s}^h$, $\mathbf{s}^v$, and $\mathbf{s}^o$ (\textit{cf.} Equation (\ref{Equ:backbone_att})). 
After that, a prediction layer is built upon the session representation $\mathbf{s}^v$ to compute both recommendation scores $\mathbf{\hat{z}}$ and recommendation probabilities $\mathbf{\hat{y}}$ (\textit{cf.} Equation (\ref{Equ:backbone_pre})). Lastly, we define the recommendation learning loss $\mathcal{L}_{rec}$ as the cross-entropy of the prediction $\mathbf{\hat{y}} \in \mathbb{R}^{|\mathcal{V}|}$\, and\, the\, ground\, truth\, $\mathbf{{y}} \in \mathbb{R}^{|\mathcal{V}|}$\, (\textit{cf.} Equation (\ref{Equ:backbone_rec})).
\subsection{Alignment and Uniformity Constraints}
As discussed in Section \ref{section1}, there exists a large gap between the \textit{attribute semantics} and \textit{collaborative semantics}, which causes a significant distribution discrepancy between the attribute-enriched item $\{\mathbf{v}_i^h\}_{i=1}^n$ and the raw item representations $\{\mathbf{v}_i^{(0)}\}_{i=1}^n$. 
Due to this, the fused item representations $\{\mathbf{v}^{sv}_{i}\}_{i=1}^{n}$ suffer from semantic indistinct and contradictory issues, which would impair the model performance. Inspired by \cite{wang2020understanding,wang2022towards,wang2021understanding,DBLP:conf/www/0001WSLZW22}, we design two representation constraints to bridge this gap: \textbf{\textit{1)}} \textbf{\textit{alignment constraint}}, which forces the representations from the same session to be as close as possible, and \textbf{\textit{2)}} \textbf{\textit{uniformity constraint}}, which forces the representations from the different sessions to be as distant as possible.
\subsubsection{\textbf{Alignment Constraint}} Considering the existence of the large gap between the attribute semantics and the collaborative semantics, it is necessary to conduct the alignment between them so that the resultant session representation $\mathbf{s}^v$ becomes more semantically accurate. 
To this end, we design a representation alignment constraint to align attribute semantics and collaborative semantics for improving the effectiveness of MIA. Specifically, it aims to minimize the distance between representations of the same session derived from different channels, whose procedure can be formulated as follows: 
\begin{equation}
    \mathcal{L}_{align} =  \underset{s \sim \mathcal{S}}{\mathbb{E}} \| \mathbf{\Tilde{s}}^h_s -  \mathbf{\Tilde{s}}^o_s \|^2_2,
    \label{Equ:alignment}
\end{equation}
where $\mathcal{S}$ is the set of the training sessions; $\mathbf{\Tilde{s}}^h_s$ and $ \mathbf{\Tilde{s}}^o_s$ are the $l_2$ normalized session representations of $\mathbf{{s}}^h_s$ and $\mathbf{{s}}^o_s$, respectively. 
\subsubsection{\textbf{Uniformity Constraint}} However, only considering the representation alignment is insufficient since the encoder is easily caught in the trivial solution by mapping all the session embeddings to the same representation. 
Therefore, it is necessary to 
conduct the alignment while preserving better uniformity so that the resultant session representation $\mathbf{s}^v$ becomes more semantically discriminative. Toward this end, we design a representation uniformity constraint, which aims to minimize the similarity between representations belong- \quad \quad ing to different session channels. It can be defined as follows:
\begin{align}
\nonumber
    \mathcal{L}_{uniform} = &\big(\log \underset{s,s^{\prime} \sim \mathcal{S}}{\mathbb{E}} e^{-2\| \mathbf{\Tilde{s}}^h_s -  \mathbf{\Tilde{s}}^h_{s^{\prime}} \|^2_2}\big)/2 + \\ &\big(\log \underset{s,s^{\prime} \sim \mathcal{S}}{\mathbb{E}} e^{-2\| \mathbf{\Tilde{s}}^o_s -  \mathbf{\Tilde{s}}^o_{s^{\prime}} \|^2_2}\big)/2,
    \label{Equ:uniformity}
\end{align}
where $\mathbf{\Tilde{s}}^h_s$ and $\mathbf{\Tilde{s}}^h_{s^{\prime}}$ are the session representations learned from attribute semantics; $\mathbf{\Tilde{s}}^o_s$ and $\mathbf{\Tilde{s}}^o_{s^{\prime}}$ are the session representations learned from collaborative semantics. 
Note that we separately calculate the uniformity constraint within each other since the distribution of attribute semantics and collaborative semantics are diverse which is more suitable to be measured respectively.
Under the alignment $\mathcal{L}_{align}$ and uniformity $\mathcal{L}_{uniform}$ constraints, representations of the same session will be close to each other, and each representation will preserve as much information about the attribute/collaborative semantics as possible. Combining them yields the final representation constraint:
\begin{equation}
    \mathcal{L}_{au} = \gamma_1\mathcal{L}_{align}+\gamma_2\mathcal{L}_{uniform}, \label{Eqa:au}
\end{equation}
where $\gamma_1$ and $\gamma_2$ are hyper-parameters controlling the strengths of the alignment and uniformity constraint, respectively.
\subsection{Model Training}
The proposed AttrGAU framework is trained based on the following learning objectives including the recommendation learning loss (\textit{cf.} Equation (\ref{Equ:backbone_rec})), the cross-layer contrast regularization (\textit{cf.} Equation (\ref{Equ:ccr})), the representation constraint (\textit{cf.} Equation (\ref{Eqa:au})), and $L_2$ regularization, formulated as follows:
\begin{equation}
    \mathcal{L} = \mathcal{L}_{rec} + \lambda_1 \mathcal{L}_{ccr} + \lambda_2 \mathcal{L}_{au} + \lambda_3 \| \Theta \|_2^2,  
\end{equation}
where $\Theta$ is the set of all learnable parameters; $\lambda_1$, $\lambda_2$, and $\lambda_3$ are hyper-parameters to control the strengths of the cross-layer contrast regularization, the representation constraints, and $L_2$ regularization, respectively. 
It is worth mentioning that the proposed AttrGAU is a model-agnostic framework that can be easily applied to existing attribute-agnostic SBR models to mitigate the severe data sparsity issue caused by short sessions.
\begin{table}[t]
  \centering
  \tabcolsep=0.175cm
  \renewcommand\arraystretch{1.25}
  \caption{\centering{Statistics of the datasets after preprocessing.}}
  \begin{tabular}{cccc}
    \toprule
    \textbf{Dataset} & \textbf{Dressipi} & \textbf{Diginetica} & \textbf{Retailrocket}\\
    \midrule 
    \#training sessions & 691,198 & 719,470 &   710,651\\
    \#test sessions & 71,272 & 60,858 &  50,095\\
    \#items & 19,728 & 43,097 &  48,929\\
    \#parent attrs & 72 & 1 &  55\\
    \#leaf attrs & 821 & 995 &  849\\
    avg.len & 6.52 & 5.12 &  5.81 \\
    \bottomrule
  \end{tabular}
  \label{tab:dataset}
\end{table}
\section{Experiments}
In this section, we conduct extensive experiments on three benchmark datasets to answer the following Research Questions (\textbf{RQs}): \textbf{RQ1:} How much can existing attribute-agnostic models gain when integrating with our proposed AttrGAU framework? \textbf{RQ2:} Can AttrGAU endow existing attribute-agnostic models with more robustness against the data sparsity problem? \textbf{RQ3:} How do different components of AttrGAU (\textit{i.e.,} the cross-layer contrast regularization and the representation constraints) contribute to final performance improvement? \textbf{RQ4:} How do different settings (\textit{e.g.,} depth of convolution layer)\, influence\, the\, effectiveness of the proposed AttrGAU?

\subsection{Experimental Settings}

\begin{table*}[th]
\centering
  \renewcommand\arraystretch{1.3}
  \tabcolsep=0.215cm
  \caption{Performance comparison of three backbone models and their AttrGAU-enhanced ones on three benchmark datasets. The `Gain' denotes the performance gain of X+AttrGAU over the vanilla X model. We use SR-GNN+, GC-SAN+, and TAGNN+ to represent AttrGAU-enhanced models for simplicity. All improvements are significant with $p$-value $<$ 0.01 based on $t$-tests.}
  \begin{tabular}{cccccc|cccccc}
    \toprule
    \multicolumn{1}{c}{\multirow{2}{*}{\centering \bf Datasets}} 
    & \multicolumn{1}{c}{\multirow{2}{*}{ \bf @N}}
    & \multicolumn{1}{c}{\multirow{2}{*}{\centering \bf Metrics}}
    & \multicolumn{3}{c|}{\multirow{1}{*}{\centering \bf Backbones}}
    & \multicolumn{6}{c}{\multirow{1}{*}{\centering \bf X+AttrGAU}} \\
     \cline{4-12}
    & & & SR-GNN & GC-SAN & TAGNN & SR-GNN+ & {\bf Gain} & GC-SAN+ & {\bf Gain} & TAGNN+ & {\bf Gain} \\
    \cline{1-12}
    \multicolumn{1}{c}{\multirow{4}{*}{\centering Dressipi}} & \multicolumn{1}{c}{\multirow{2}{*}{\centering @5}} & HR & 25.25 & 23.41 & 25.63 & 26.65 & {\bf 5.54\%} & 25.05 & {\bf 7.01\%} &  27.37 & {\bf 6.79\%} \\
     &  & MRR & 16.32 & 13.58 & 15.83 & 17.56 & {\bf 7.60\%} & 14.39 & {\bf 5.96\%} & 18.30 & {\bf 15.6\%}\\
      \cline{2-2}
     & \multicolumn{1}{c}{\multirow{2}{*}{\centering @10}} & HR & 32.19 & 31.27 & 33.40 &  33.55 & {\bf 4.22\%} &  32.90 & {\bf 5.21\%} &  34.10 & {\bf 2.10\%}\\
     & & MRR & 17.25 & 14.63 & 16.87 & 18.48 & {\bf 7.13\%} & 15.45 & {\bf 5.60\%} & 19.20 & {\bf 13.8\%}\\

    \cline{1-12}
    \multicolumn{1}{c}{\multirow{4}{*}{\centering Diginetica}} & \multicolumn{1}{c}{\multirow{2}{*}{\centering @5}} & HR & 26.16 & 25.30 & 26.61 & 27.53 & \textbf{5.24\%} & 26.98 & \textbf{6.64\%} &  27.80 & \textbf{4.47\%} \\
     &  & MRR & 14.49 & 14.18 & 14.99 & 15.44  & \textbf{6.56\%} & 15.38 & \textbf{8.46\%} & 15.86 & \textbf{5.80\%} \\
      \cline{2-2}
     & \multicolumn{1}{c}{\multirow{2}{*}{\centering @10}} & HR & 36.93  & 36.41 & 37.48 & 38.59 & \textbf{4.49\%} & 38.05 & \textbf{4.50\%} & 39.01 &  \textbf{4.08\%}\\
     &  & MRR & 16.22 & 15.95 & 16.51 & 16.93 & \textbf{4.38\%} & 16.85 & \textbf{5.64\%} & 17.36 & \textbf{5.15\%}\\

     \cline{1-12}
    \multicolumn{1}{c}{\multirow{4}{*}{\centering Retailrocket}} & \multicolumn{1}{c}{\multirow{2}{*}{\centering @5}} & HR & 48.35 & 44.87 & 48.73 &  49.69 & {\bf 2.77\%} & 48.13 & \textbf{7.27\%} &  49.91 & \textbf{2.42\%} \\
     &  & MRR & 34.16 & 31.97 & 34.46 & 35.65 & {\bf 4.36\%} & 34.81 & \textbf{8.88\%} & 36.07 & \textbf{4.67\%} \\
      \cline{2-2}
     & \multicolumn{1}{c}{\multirow{2}{*}{\centering @10}} & HR & 56.93 & 52.86 & 57.25 & 58.27 & {\bf 2.35\%} & 56.20 & \textbf{6.32\%} &  58.32 & \textbf{1.87\%} \\
     &  & MRR & 35.58 & 33.04 & 35.61 & 36.81 & \textbf{3.46\%} & 35.89 & \textbf{8.63\%} & 37.20 & \textbf{4.47\%} \\
     
    \bottomrule
  \end{tabular}
  \label{tab:overall}
\end{table*}
\begin{table*}[ht]
  \renewcommand\arraystretch{1.25}
  \tabcolsep=0.28cm
  \caption{Performance comparison \textit{w.r.t.} different percentage of training data (\%) on three benchmark datasets. The percentage in brackets denotes the relative performance improvement over its corresponding vanilla ones, \textit{e.g.,} SR-GNN+ versus SR-GNN.}
  \centering
  \begin{tabular}{c|l|c|c|c|c|c|c}
    \toprule
    \multicolumn{2}{c|}{ \multirow{1}{*}[+0.2ex]{ \centering \bf Datasets}} & \multicolumn{2}{c|}{ \multirow{1}{*}[+0.2ex]{ \centering \bf Dressipi}} &  \multicolumn{2}{c|}{ \multirow{1}{*}[+0.2ex]{\centering \bf Diginetica}} & \multicolumn{2}{c}{ \multirow{1}{*}[+0.2ex]{ \centering \bf Retailrocket}}  \\
    \cline{1-8}
     \multicolumn{1}{c|}{ \multirow{1}{*}[-0.2ex]{ \bf Percentage}} & \multicolumn{1}{c|}{ \multirow{1}{*}[-0.2ex]{ \bf Models}} & \multicolumn{1}{c|}{ \multirow{1}{*}[-0.2ex]{ \bf HR@5}} & \multicolumn{1}{c|}{ \multirow{1}{*}[-0.2ex]{ \bf MRR@5}} & \multicolumn{1}{c|}{ \multirow{1}{*}[-0.2ex]{ \bf HR@5}} & \multicolumn{1}{c|}{ \multirow{1}{*}[-0.2ex]{ \bf MRR@5}} & \multicolumn{1}{c|}{ \multirow{1}{*}[-0.2ex]{ \bf HR@5}} & \multicolumn{1}{c}{ \multirow{1}{*}[-0.2ex]{ \bf MRR@5}}  \\
     
      \cline{1-8}
      \multicolumn{1}{c|}{ \multirow{6}{*} { \bf 25 Percent}} & SR-GNN & 13.54 & 7.984  & 18.67 & 10.60 & 37.26 & 26.96 \\
      & GC-SAN & 13.11 & 7.393 & 15.70 & 8.709 & 28.30 & 20.32 \\
      & TAGNN & 13.84 & 7.998 & 18.90 & 10.86 & 39.50 & 28.34 \\
      & \textbf{SR-GNN+} & \textbf{18.55(37.0\%)} & \textbf{11.68(46.3\%)} & \textbf{22.34(19.7\%)} & \textbf{12.92(21.9\%)} & \textbf{42.29(13.5\%)} & \textbf{31.10(15.4\%)} \\
      & \textbf{GC-SAN+} & \textbf{15.07(15.0\%)} & \textbf{8.218(11.2\%)} & \textbf{18.73(19.3\%)} & \textbf{10.97(26.0\%)} & \textbf{35.46(25.3\%)} & \textbf{26.50(30.4\%)}\\
      & \textbf{TAGNN+} & \textbf{19.76(42.8\%)} & \textbf{12.64(58.0\%)} & \textbf{22.96(21.5\%)} & \textbf{13.49(24.2\%)} &  \textbf{42.56(7.75\%)} & \textbf{31.57(11.4\%)} \\

    \cline{1-8}
     \multicolumn{1}{c|}{ \multirow{6}{*} { \bf 50 Percent}} & SR-GNN & 20.49 & 12.21 & 21.67 & 12.15 & 42.71 & 30.13 \\
      & GC-SAN & 18.91 & 10.81 & 20.55 & 11.26 & 39.92 & 28.70\\
      & TAGNN & 21.50 & 12.53 & 22.58 & 12.45 & 43.95 & 31.27 \\
      & \textbf{SR-GNN+} & \textbf{23.54(14.9\%)} & \textbf{15.19(24.4\%)} & \textbf{23.18(6.97\%)}  & \textbf{13.44(10.6\%)} & \textbf{46.53(8.94\%)} & \textbf{33.74(12.0\%)} \\
      & \textbf{GC-SAN+} & \textbf{21.76(15.1\%)} & \textbf{11.95(10.5\%)} & \textbf{23.36(13.7\%)} & \textbf{13.39(18.9\%)} & \textbf{43.61(9.24\%)} & \textbf{31.94(11.3\%)}\\
      & \textbf{TAGNN+} & \textbf{24.03(11.8\%)} & \textbf{15.65(24.9\%)} & \textbf{24.43(8.19\%)} & \textbf{14.27(14.6\%)} & \textbf{46.33(5.42\%)} & \textbf{33.78(8.03\%)} \\

      \cline{1-8}
     \multicolumn{1}{c|}{ \multirow{6}{*} { \bf 75 Percent}} & SR-GNN & 22.97 & 14.41 & 23.96 & 13.89 & 45.80 & 31.67 \\
      & GC-SAN & 22.18 & 12.64 & 23.08 & 12.71 & 43.43 & 32.26\\
      & TAGNN & 24.07 & 14.49 & 23.88 & 13.95 & 45.36 & 31.86\\
      & \textbf{SR-GNN+} & \textbf{25.38(10.5\%)} & \textbf{16.66(15.6\%)} & \textbf{26.03(8.64\%)} & \textbf{15.13(8.93\%)} & \textbf{48.52(5.94\%)} & \textbf{34.94(10.3\%)}\\
      & \textbf{GC-SAN+} & \textbf{23.70(6.85\%)} & \textbf{13.17(4.19\%)} & \textbf{25.73(11.5\%)} & \textbf{14.70(15.7\%)} & \textbf{46.45(6.95\%)} & \textbf{33.72(4.53\%)}\\
      & \textbf{TAGNN+} & \textbf{26.05(8.23\%)} & \textbf{17.37(19.9\%)} & \textbf{26.45(10.8\%)} & \textbf{15.19(8.89\%)} & \textbf{48.03(5.89\%)} & \textbf{34.74(9.04\%)}\\
    \bottomrule
  \end{tabular}
  \label{tab:percentage}
\end{table*}
\subsubsection{\textbf{Datasets and Preprocessing} \label{sec:datasets}} We adopt three public benchmark datasets to evaluate our framework, \textit{i.e.,} Dressipi\footnote{https://www.recsyschallenge.com/2022/dataset.html}, Diginetica\footnote{https://competitions.codalab.org/competitions/11161}, Retailrocket\footnote{https://www.kaggle.com/datasets/retailrocket/ecommerce-dataset}. 
Particularly, the Dressipi dataset is from RecSys Challenge 2022, consisting of viewed and purchased logs. 
The Diginetica dataset comes from CIKM Cup 2016, containing anonymized search and browsing logs. The Retailrocket dataset is released by a personalized e-commerce company, which is composed of user browsing logs. Following \cite{xu2019graph,wu2019session,xia2021self}, we filter out sessions of length 1 and items appearing less than 5 times across all three benchmark datasets, where the behaviors with the same session identifier are treated as a session directly in the Dressipi and Diginetica datasets, and the continuous user behaviors within 30 minutes are treated as a session in the Retailrocket dataset. 
Similar to \cite{liu2018stamp,zheng2022heterogeneous,su2023enhancing}, we set the sessions of the most recent ones (\textit{i.e.,} one month for Dressipi, one week for Diginetica, and two days for Retailrocket) as the test data, and the remaining for training data. After that, given a session $s = \{v_{s,1}, v_{s,2}, ...,v_{s,n}\}$ from training or test set, we generate behavior sequences and corresponding labels by a sequence splitting process across all the three datasets, \textit{i.e.,} $([v_{s,1}], v_{s,2}), ([v_{s,1},v_{s,2}], v_{s,3}),...,([v_{s,1}, v_{s,2}, ...,v_{s,n-1}], v_{s,n})$. The detailed statistics of three public benchmark datasets after\, preprocessing\, procedures\, are\, summarized\, in\, Table\, \ref{tab:dataset}.
\subsubsection{\textbf{Evaluation Metrics}} We adopt two commonly used metrics for performance evaluation, \textit{i.e.,} Hit Rate (HR@N) and Mean Reciprocal Rank (MRR@N). Specifically, the former measures the proportion of the ground truth item in an unranked list, while the latter further considers the position of the ground truth item in a ranked list. And the larger the values the better the recommendation performance for both of them. In our experiments, we report the results of $\mathrm{N}=5,10$. 
\subsubsection{\textbf{Backbone Models}} The proposed AttrGAU is a model-agnostic MIA framework that aims to enhance the recommendation performance as well as the resistance to data sparsity of existing attribute-agnostic SBR models. To verify whether AttrGAU can achieve the above goals, we test its capacity based on the following three\, representative\, SBR\, backbones:
\begin{itemize}
    \item \textbf{SR-GNN} \cite{wu2019session}. It adopts a gated GNN layer to obtain item embeddings by modeling contextual transitions and then generates the\, session\, embedding\, via\, additive\, attention.
    \item \textbf{GC-SAN} \cite{xu2019graph}. Like SR-GNN, it also refines the item embeddings via a gated GNN layer but learns to generate a more comprehensive session embedding by stacking multiple self-attention layers instead of additive attention.
    \item \textbf{TAGNN} \cite{yu2020tagnn}. Different from GC-SAN which mines users' comprehensive interests, it adaptively extracts users' \quad \quad diverse\, interests\, in\, sessions via a target attentive layer. 
\end{itemize}
\subsubsection{\textbf{Implementation Details}} For a fair comparison, we adopt the same hyper-parameter settings as those reported in their released source codes. Specifically, we set the hidden dimensionality as 100 for SR-GNN and TAGNN and 120 for GC-SAN. We use the Adam optimizer \cite{kingma2014adam} to optimize model parameters with the learning rate of 0.001, the mini-batch size of 100, $\beta_1$=0.9, and $\beta_2$=0.999, where the learning rate will decay by 0.1 after every 3 epochs. Moreover, the maximum number of epochs is set to 30, and $L_2$ regularization coefficient $\lambda_3$ is set to 1$e^{-5}$. During training, we adopt early stopping on the test set if the performance does not improve for 10 epochs. 
We implement AttrGAU in PyTorch \cite{paszke2019pytorch} and employ a grid search to find the proper hyper-parameters. Specifically, we tune the number of graph convolution layers $L$ within $\{1,\mathbf{2},\mathbf{3},4\}$; for the weight coefficient of each learning objective, we tune $\gamma_1$, $\gamma_2$, $\lambda_1$, and $\lambda_2$ within the range of $\{\mathbf{0.25}, 0.5, 0.75, 1.0\}$, $\{0.1, 0.2, 0.5, \mathbf{1.0}\}$, $\{\mathbf{0.0005}, 0.001, 0.005, 0.01, 0.05, 0.1\}$, and $\{0.1, 0.2, 0.3, ..., \mathbf{1.0}\}$, respectively, where the boldfaced ones are favorable setting during training. Besides, we set the temperature coefficient $\tau$ in the cross-layer contrast regularization\, as\, 0.2\, and train all SBR models from scratch. 
\subsection{Model-agnostic Gain (RQ1)}
We train all SBR backbone models and their AttrGAU-enhanced ones on three benchmark public datasets. From the experimental results shown in Table \ref{tab:overall}, we mainly have the following three observations: \textbf{(1)} AttrGAU-enhanced models consistently and substantially perform better than their corresponding vanilla ones on all three datasets in terms of all metrics. For example, the average improvements of AttrGAU-enhanced models over their corresponding vanilla ones on the three datasets are 5.35\% and 7.54\% in terms of HR@5 and MRR@5. 
\textbf{(2)} AttrGAU-enhanced models only introduce a small amount of additional trainable parameters compared with the vanilla ones, \textit{i.e.,} the embedding matrix of parent attr and leaf attr, which shows the proposed AttrGAU is memory efficient and lightweight. \textbf{(3)} Since AttrGAU and its backbone model share the same graph neural network sub-module and the attention readout sub-module, it demonstrates that the proposed AttrGAU is not only good at bringing the MIA's superiority into existing attribute-agnostic SBR backbone models but also satisfies the non-intrusive requirement (\textit{cf.} Section \ref{section1}). 
\begin{table}[t]
  \renewcommand\arraystretch{1.3}
  \tabcolsep=0.225cm
  \caption{Ablation study with key components, where the best results are boldfaced and the worst results are underlined. Here, we use HR and MRR to indicate HR@5 and MRR@5.}
  \centering
  \begin{tabular}{c|l|cc|cc}
    \toprule
    \multicolumn{2}{c|}{ \multirow{1}{*}[+0.2ex]{\centering \bf Datasets}} & 
    \multicolumn{2}{c|}{ \multirow{1}{*}[+0.2ex]{\centering \bf Dressipi}} & 
    \multicolumn{2}{c}{ \multirow{1}{*}[+0.2ex]{\centering \bf Diginetica}} \\
    \cline{1-6}
     \multicolumn{1}{c|}{ \multirow{1}{*}[-0.2ex]{\centering \bf Models}} & \multicolumn{1}{c|}{ \multirow{1}{*}[-0.2ex]{\centering \bf Variants}} &\multirow{1}{*}[-0.2ex]{\centering \bf HR} & \multirow{1}{*}[-0.2ex]{\centering \bf MRR} & \multirow{1}{*}[-0.2ex]{\centering \bf HR} & \multirow{1}{*}[-0.2ex]{\centering \bf MRR}  \\
     
    \cline{1-6}
     \multicolumn{1}{c|}{ \multirow{4}{*} {\centering \bf SR-GNN+}} & \textbf{(A) Full} & \textbf{26.65} & \textbf{17.56} & \textbf{27.53} & \textbf{15.44}  \\
      & (B) w/o $\mathcal{L}_{ccr}$ & 25.82 & 16.88 & {27.19} & 15.19\\
      & (C) w/o $\mathcal{L}_{align}$ & {26.37} & {17.06} & 26.98 & {15.27} \\
      & (D) w/o $\mathcal{L}_{uniform}$ & \underline{23.87} & \underline{14.54} & \underline{26.84} & \underline{15.06}\\

      \cline{1-6}
     \multicolumn{1}{c|}{ \multirow{4}{*} {\centering \bf GC-SAN+}} & \textbf{(A) Full}& \textbf{25.05} & \textbf{14.39} & \textbf{26.98} & \textbf{15.38} \\
      & (B) w/o $\mathcal{L}_{ccr}$ & \underline{23.95} & \underline{13.58} & 26.35 & 14.76\\
      & (C) w/o $\mathcal{L}_{align}$ & 24.17 & 13.45 & {26.69} & {15.16} \\
      & (D) w/o $\mathcal{L}_{uniform}$ & {24.44} & {13.78} & \underline{26.15} & \underline{14.55} \\

      \cline{1-6}
     \multicolumn{1}{c|}{ \multirow{4}{*} {\centering \bf TAGNN+}} & \textbf{(A) Full} & \textbf{27.37} & \textbf{18.30} & \textbf{27.80} & \textbf{15.86} \\
      & (B) w/o $\mathcal{L}_{ccr}$ & 26.90 & {17.86} & \underline{27.07} & \underline{15.41} \\
      & (C) w/o $\mathcal{L}_{align}$ & {27.08} & 17.67 & {27.39} & {15.69} \\
      & (D) w/o $\mathcal{L}_{uniform}$ & \underline{25.64} &  \underline{15.86} & {27.24} & {15.47} \\
    \bottomrule
  \end{tabular}
  \label{tab:ablation}
\end{table}
\begin{figure}[t]
    \subfigure[Dressipi dataset.]{
        \includegraphics[width=1.0\linewidth]{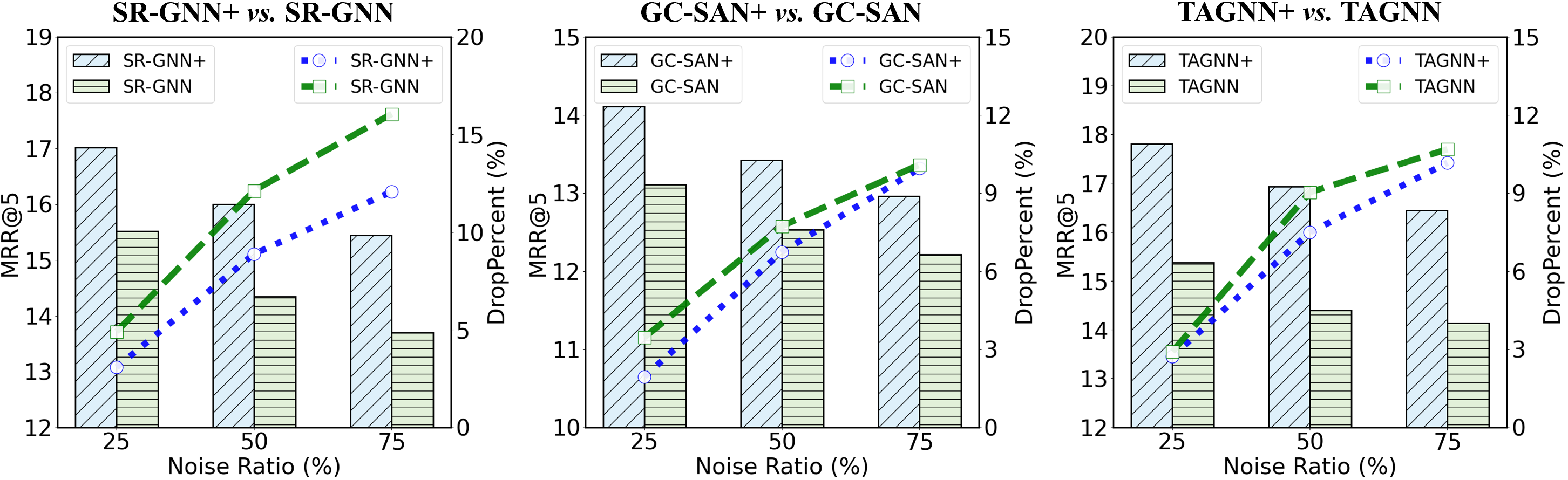}\label{fig:noise_dress}}
    \subfigure[Diginetica dataset.]{
        \includegraphics[width=1.0\linewidth]{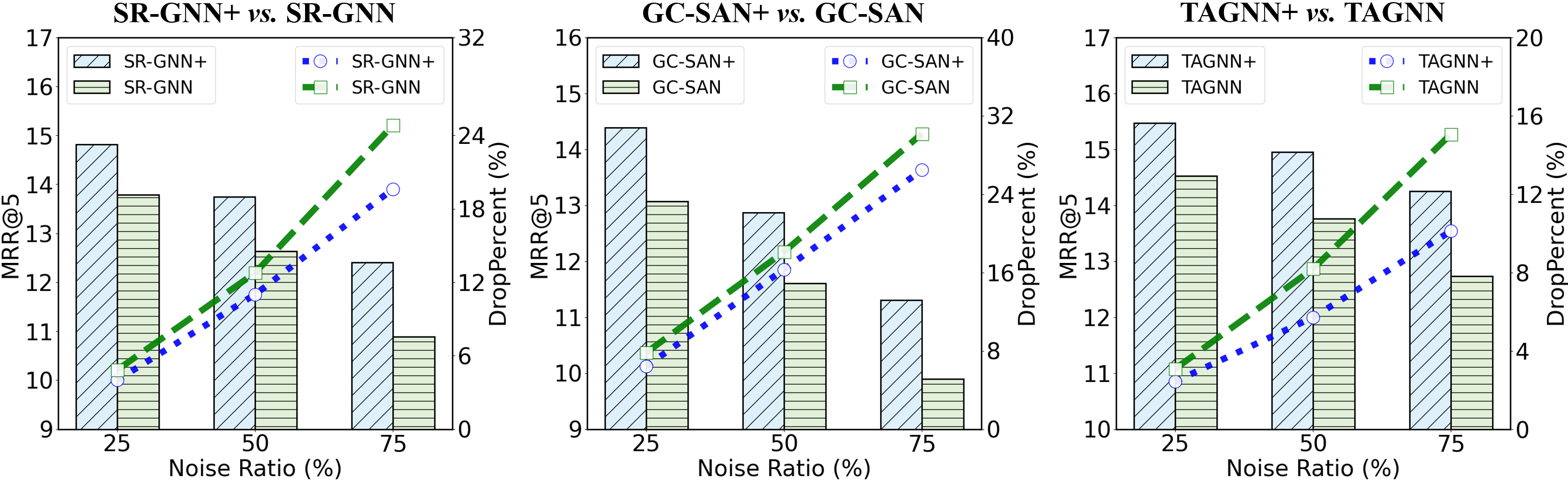}\label{fig:noise_dig}}
    \caption{Model performance \textit{w.r.t.} noise ratio on Dressipi and Diginetica datasets.  The bar represents MRR@5, while the line represents the percentage of performance degradation compared\, with\, the\, model\, trained on 100\% of training data.}
    \label{fig:robust_noise}
\end{figure}
\begin{figure*}[t]
	\centering  
	\subfigure[Effect of model depth $L$ on two datasets.]{
		\includegraphics[width=0.37\linewidth]{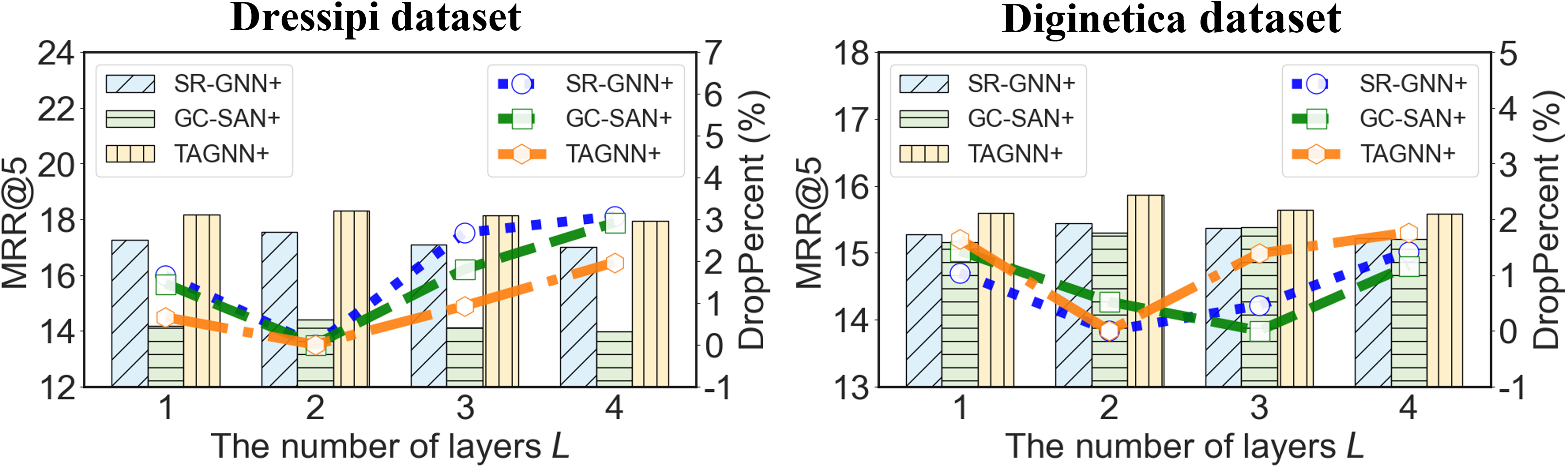}\label{fig:layer_hyper}}
	\subfigure[Performances \textit{w.r.t.} different groups on Dressipi. The line denotes the relative improvement.]{
		\includegraphics[width=0.61\linewidth]{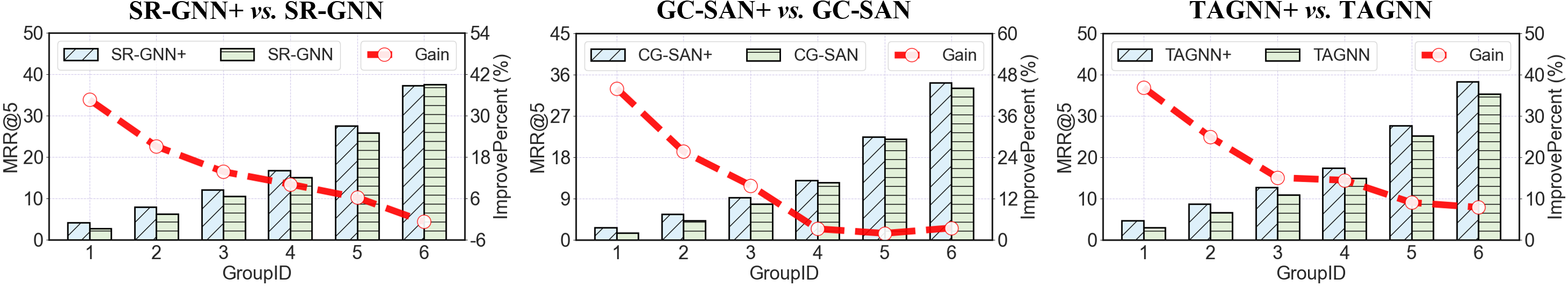}\label{fig:popu_analy}}
	\caption{Sensitivity study on model depth $L$ (1, 2, 3, and 4) and performances \textit{w.r.t.} item popularity (1, 2, 3, 4, 5, and 6).}
    \label{fig:layer_popu_analy}
\end{figure*}
\subsection{Robustness to Sparse Data (RQ2) \label{robust_sparse}}
In SBR scenarios, a prevalent challenge is the data sparsity issue caused by short sessions, \textit{e.g.,} the average session length in Dressipi, Diginetica, and Retailrocket datasets is 6.52, 5.12, and 5.81, respectively, which inflicts a heavy blow to the performance of SBR models. To study the robustness of AttrGAU against sparse data, we train models with only partial training data (\textit{i.e.,} 25\%, 50\%, and 75\%) and keep the test data unchanged. Table \ref{tab:percentage} shows the results on three datasets. We find that: \textbf{(1)} Model performance substantially degrades when using less training data, but AttrGAU-enhanced models consistently outperform their corresponding vanilla ones. 
For example, they achieve comparable performance with only 75\% of training data as of the vanilla ones with 100\% of training data. \textbf{(2)} The more sparse the training data, the greater the performance improvement. For example, when using 75\% of the training data, SR-GNN+ improves by 15.6\% compared to SR-GNN, while using 25\% of the data, the improvement is 46.3\%, on MRR@5. These observations show that the proposed\, AttrGAU\, can mitigate the data sparsity issue well. 
\subsection{Ablation Study (RQ3)}
To assess the effectiveness of individual components within our framework, we conduct several ablation experiments on AttrGAU by removing $\mathcal{L}_{ccr}$, $\mathcal{L}_{align}$, and $\mathcal{L}_{uniform}$, respectively. Table \ref{tab:ablation} summarizes the overall performances of different variants, where the `\textbf{Full}' means the complete version. \textbf{Firstly}, from the table, we can find that the `\textbf{Full}' achieves the best results on all datasets, which indicates all components are effective and necessary for our framework.
\textbf{Secondly}, by comparing (B), (C), and (D), we observe that removing $\mathcal{L}_{uniform}$ generally results in the greatest performance degradation and removing $\mathcal{L}_{align}$ also leads to large performance degradation, which suggests enforcing session representations to be more discriminative by the uniformity constraint is of significance and bridging the gap between the \textit{attribute semantics} and \textit{collaborative semantics} is valuable. \textbf{Thirdly}, by comparing (A) and (B), it can be observed that mitigating the over-smoothing issue by the cross-layer contrast regu- \quad\quad larization could significantly improve the model performance.
\subsection{Study of AttrGAU (RQ4)}
We move on to studying different settings in the proposed AttrGAU. We \textbf{first} assess the robustness of AttrGAU against noisy data. We \textbf{then} investigate the impact of the graph convolution layer $L$. We \textbf{finally} explore the potential capacity of our\, AttrGAU\, to\, enhance\, the\, long-tail\, recommendation.
\subsubsection{\textbf{Robustness to Noisy Data}} As shown in Figure \ref{fig:robust_noise}, we experiment to verify AttrGAU's robustness against noisy data. Specifically, we train models with full training data but randomly add a certain ratio (\textit{i.e.,} 25\%, 50\%, and 75\%) of negative items into test sessions. 
From the experimental results, we observe that adding noisy interactions significantly degrades the performance of AttrGAU-enhanced models and their corresponding vanilla ones. However, the performance degradation of AttrGAU-enhanced models is always lower than their vanilla ones. This shows that AttrGAU can figure out useful semantic patterns and endows the backbone with more robustness against noisy data during the inference stage.
\subsubsection{\textbf{Impact of Model depth}} As shown in Figure \ref{fig:layer_hyper}, we experiment to investigate the impact of model depth, where we search the number of AttrGC layers $L$ in the range of $\{1,2,3,4\}$. It can be observed that AttrGAU-enhanced models get a peak value at medium depth, which manifests the effectiveness of the proposed AttrGC and manifests that setting a suitable number of AttrGC layers can boost model performance. 
Specifically, $L=2$ for Dressipi and $L=2,3$ for Diginetica are generally appropriate to AttrGAU, consistent with our statement that over-smoothing by excessive graph\, convolution\, will\, inevitably\, cause\, performance\, degradation. 
\subsubsection{\textbf{Long-tail Recommendation}} As shown in Figure \ref{fig:popu_analy}, we experiment to verify whether AttrGAU can enhance long-tail recommendation. Specifically, we split the test sessions into 6 groups based on the target item's popularity (the number of interactions) and ensure the number of test sessions within each group is the same, where the larger the GroupId, the more popular the target item. From the experimental results, we observe that the gain brought by AttrGAU generally decreases as the popularity of items increases. This verifies that AttrGAU can establish better representations for long-tail items and hence improve the performance of long-tail recommendations.
\section{Related Works}
\subsection{Conventional SBR Methods} 
Pioneering attempts on SBR are based on Markov chains to model item-item transition patterns and then predict the next-click item \cite{shani2005mdp,rendle2010factorizing}. However, they only take into account the most recent clicked item within the session and thus restrict the prediction accuracy. 
To model long-term dependencies, Recurrent Neural Networks (RNNs) have been widely used for SBR by modeling sequence-level item transitions \cite{hidasi2015session,tan2016improved,jannach2017recurrent,li2017neural,hidasi2018recurrent,hidasi2016parallel,DBLP:conf/sigir/WangRMCMR19}. For example, GRU4Rec \cite{hidasi2015session} employs Gated Recurrent Unit (GRU) to learn the evolving patterns within the session and generate user preference. Despite effectiveness, they only model the unidirectional transition between consecutive items, and\, fall\, short\, of\, mining\, the complex contextual transitions.
\subsection{GNN-based SBR Methods} Owing to its strong representation capabilities, GNN has been widely used to make SBR \cite{wu2019session,xu2019graph,yu2020tagnn,chen2020handling}. SR-GNN \cite{wu2019session} is the pioneering work in adopting GNN for SBR, which converts sessions into directed graphs and employs GGNN to model complex item transitions. Based on that, GC-SAN \cite{xu2019graph} applies the self-attention mechanism on the item representations learned by GGNN, to model long-range dependencies among items. 
On the other hand, TAGNN \cite{yu2020tagnn} thinks that the target items play an important role in extracting the underlying users’ interests, and propose target-aware attention to adaptively activate different user interests in terms of varied target items. Furthermore, LESSR \cite{chen2020handling} identifies the information loss issue of GNNs for SBR, and proposes edge-order preserving aggregation and shortcut graph attention to address this issue. While encouraging, these works rely too heavily on contextual transitions, which is largely limited by the data sparsity issue.
\subsection{Attribute-aware SBR Methods} Recently, a few works \cite{lai2022attribute,ma2023clhhn,xu2022category} have attempted to leverage additional exogenous knowledge, especially the attribute of items, to alleviate the data sparsity issue caused by short sessions. For example, CM-HGNN \cite{xu2022category} builds an item-category heterogeneous graph to model item-item, item-category, and category-category patterns, simultaneously. 
MGS \cite{lai2022attribute} performs interactive dual refinement on the built session graph and attribute-driven mirror graph to fuse session-wise and attribute-wise semantics. CLHHN \cite{ma2023clhhn} explicitly models the complex relations among items and categories by constructing a lossless session heterogeneous hypergraph. However, these works involve specific model designs that can hardly transfer their superiority to existing SBR models, lacking universality.
\section{CONCLUSION}
In this paper, we emphasize the importance of bringing the MIA's superiority into existing attribute-agnostic SBR models and disclose two main challenges hindering its development, \textit{i.e.,} {Heterogeneous Item-Attribute Relationship} and {Distribution Discrepancy in Representation}. 
To this end, we propose a novel attributed learning framework, AttrGAU, which extracts the rich item-attr semantics from the well-organized bipartite attributed graph (BAG) and learns to bridge the large gap between the \textit{attribute semantics} and \textit{collaborative semantics} by representation constraints. Different from a few existing attribute-aware SBR models that lack universality, the proposed AttrGAU is lightweight, model-agnostic, and flexible for plug-and-play usage. We have conducted extensive experiments on three benchmark datasets. The experimental results show that AttrGAU can effectively improve backbone models' recommendation performance, robustness against sparse and noisy data, as well as long-tail recommendation performance.
\balance
\bibliographystyle{IEEEtran}
\bibliography{IEEEabrv,ref}

\end{document}